\documentstyle[aps,twocolumn,epsf]{revtex} 

\begin{document}
\title{On the Path Integral of the Relativistic Electron}
\author{
Andreas Kull\thanks{{\it Send offprint requests to:} kull@cita.utoronto.ca}\\
Canadian Institute for Theoretical Astrophysics, 60 St. George Street, Toronto, M5S 1A7, Canada\\[0.125cm] 
and \\[0.125cm]
Rudolf A. Treumann \\
Max-Planck-Institut f\"ur Extraterrestrische Physik, D-85740 Garching, Germany}
\maketitle

\begin{abstract}
We revisit the path integral description of the motion of a relativistic electron.
Applying a minor but well motivated conceptional change to Feynman's 
chessboard model, we obtain exact solutions of the Dirac equation. 
The calculation is performed by means of a particular simple method different
from both the combinatorial approach envisaged by Feynman and its Ising model 
correspondence.
\end{abstract}

\section{Introduction}
It is well known that the continuum propagator of the Dirac equation can be found by 
summing over random walks. Renewed interest in this issue has arisen
in connection with the investigation of stochastic processes which have been 
shown to be related to the Dirac equation (Gaveau et al. 1984, McKeon and Ord 1992). 
Likewise, the correspondence between the path integral and the Ising model has been 
explored (Gersch 1981, Jacobson and Schulman 1984) and solutions for a discretized version 
of the Dirac equation have been found (Kauffman and Noyes, 1996).

As described by Feynman and Hibbs (1965), the propagator of 
the $1+1$ dimensional Dirac equation
\begin{equation}
\mbox{i} {{\partial \Psi}/{\partial t}}=
-\mbox{i} \sigma_z {{\partial \Psi}/{\partial x}} - m \sigma_x \Psi
\label{diracequ}
\end{equation} 
(where units $c=\hbar=1$ are assumed and $\sigma_x$ and $\sigma_z$ 
are the respective Pauli spin matrices) can be found from a model of the 
one-dimensional motion of a relativistic particle. In this model the motion of
the electron is restricted to movements either forward or backward occuring at
the speed of light. Assuming units $c=1$, the motion of the particle
corresponds to a sequence of straight path segments of slope $\pm 45^\circ$ in
the x-t plane. The retarded propagator $K(x,t)$ of the Dirac equation may then 
obtained from the limiting process (see e.g. Feynman and Hibbs 1965, 
Jacobson and Schulman 1984)
\begin{equation} 
K_{\delta\gamma}(x,t) = \lim_{N \rightarrow \infty} A_{\delta\gamma}(\epsilon)
\sum_{R \ge 0} N_{\delta\gamma}(R) (\mbox{i} m \epsilon)^{R} \;.
\label{proplin}
\end{equation}   

Here, $N$ is the number of segments of constant length $\epsilon = t/N$ of the
particle's path between its start point (which is assumed to be the origin of
the corresponding coordinate system) and the end point $(x,t)$ of the path. 
$R$ denotes the number of bends while $N_{\delta\gamma}(R)$ stands for the
total number of paths consisting of $N$ segments with $R$ bends. The indices
$\gamma$ and $\delta$ correspond to the directions forward or backward at the
path's start and end points, respectively, and refer to the components of
$K$. $A_{\delta\gamma}(\epsilon)$ accounts for a convenient normalization.

\section{Model and Calculations}
\begin{figure} 
\begin{center}
\leavevmode{\epsfxsize=7.5cm\epsffile{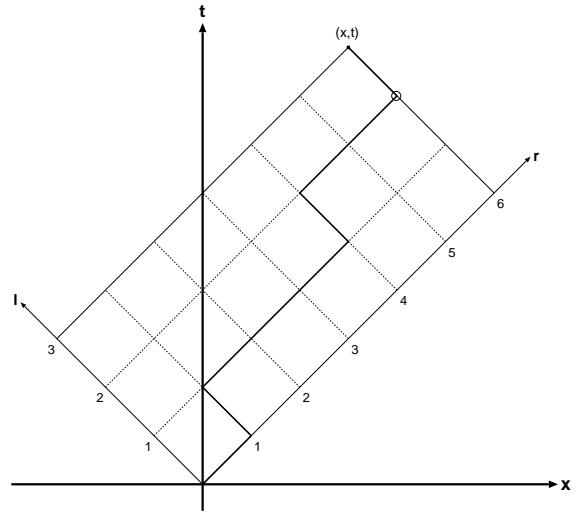}}
\end{center}
\caption{A possible path with five bends between given start and end points. 
The first two bends to the right and left, respectively, determine the
path since the location and direction of the last bend (indicated by a circle)
is fully determined by the first four bends.\label{fig1}}
\end{figure}

In this short note we demonstrate that a minor conceptional change of Feynman's
chessboard model naturally and directly yields {\it exact} solutions to the 
Dirac equation (\ref{diracequ}).
The conceptional change is suggested by the observation that a path with $R$ 
bends between given start and end points is determined by $R-1$ bends. For a
sketch of the situation consider Figure \ref{fig1}. The path shown in Figure
\ref{fig1} exhibits five bends, three to the left and two to the right.
However, the first two bends to the right and left, respectively, determine the
path since the location and direction of the last bend (indicated by a circle
in Figure \ref{fig1}) is fully determined by the first four bends. We thus
consider here, in contrast to the original formulation of the model where all
bends occuring on a path contribute to the total amplitude, only contributions
to the total amplitude from bends which {\it actually} define the path. In the
light of the general path integral formalism it makes perfect sense to
consider only those bends which define a path, i.e. the {\it minimum}
information characterizing a path.

In the following we demonstrate by an explicit calculation that the modified
model directly leads to exact solutions of the Dirac equation (\ref{diracequ}).
We will use a calculation scheme different from  the combinatorial approach 
envisaged by Feynman and Hibbs (1965) and its Ising model correspondence
(Gersch, 1981). Following Feynman's chessboard model we consider each bend
which defines a possible path to contribute an amplitude
\begin{equation}
\phi_{j_r}=\mbox{i} m \epsilon_{j_r}
\label{alin1}
\end{equation} 
where $\epsilon_{j_r}\doteq \epsilon$ is the length of a path segment. The 
total amplitude contributed by a path is the product
\begin{equation}
\phi=\prod^{}_{r} (\mbox{i} m \epsilon_{j_r})
\label{alin2}
\end{equation} 
where $j_r$ runs over all the segments followed by a bend. While the index 
$r$ enumerates the path segments after which bends occur, the value of $j_r$
indicates the corresponding segment.
A path with $R$ bends which starts with positive velocity (i.e. to the right)
and ends with negative velocity (i.e. to the left) consists of exactly
$(R-1)/2+1$ bends to the left and $(R-1)/2$ to the right. The $(R-1)/2$ bends
to the right may occur after any arbitrary path segment to the left. $(R-1)/2$
of the $(R-1)/2+1$ bends to the left occur in the same manner after path
segments to the right while the additional bend to the left occurs after the
last segment. Let $P$ be the total number of path segments to the right (+) and
$Q$ those to the left (-). Then, the contribution of the  $R^+=(R-1)/2$ bends
to the right to $\Psi_{-+}$ is 
\begin{eqnarray}
\Psi_{-+}(R^+) &=& N_{-+}(R^+)\prod^{R^+}_{r=1} (\mbox{i} m \epsilon_{j_r}) \nonumber \\
 	    &=& \sum^{P-1}_{j_1 < \ldots < j_{R^+}} (\mbox{i} m \epsilon_{})^{R^+}
\end{eqnarray}	 
For $P \gg 1$, $\Psi_{-+}(R^+)$ is approximated by
\begin{eqnarray}
\Psi_{-+}(R^+) &\approx& {1\over{{R^+}!}}\sum^{P}_{j_1 \neq\ldots\neq j_{R^+}} (\mbox{i}\epsilon_{})^{R^+} \nonumber \\ 
 	    &\approx& {(\mbox{i} m \epsilon_{})^{R^+}\over{{R^+}!}}\left( \sum_{j_r=1}^{P} 1 \right) ^{R^+} \nonumber \\
 	    &=& { P^{R^+} (\mbox{i} m \epsilon_{})^{R^+}\over{{R^+}!}}
\end{eqnarray}	
The contribution of the $R^-=[((R-1)/2+1)-1]=(R-1)/2$ bends to the left is
calculated similarly. The additional bend (occuring after the last
segment to the right) does not enter the calculation since a possible path is
fully determined by the location by its $R-1$ bends to the right and left, 
respectively. Therefore we find 
\begin{eqnarray}
\Psi_{-+}(R^-) &\approx& { Q^{R^-} (\mbox{i}\epsilon_{})^{R^-}\over{{R^-}!}}
\end{eqnarray}	
In the limit $N \rightarrow \infty$ (i.e. $P,Q \rightarrow \infty$) the exact 
expression for $\Psi_{-+}$ becomes
\begin{equation}
\Psi_{-+}=\sum_{\mbox{\tiny odd } R} (\mbox{i} m \epsilon_{})^{R-1}
{(PQ)^{(R-1)/2}\over{[((R-1)/2)!]^2}}
\end{equation}
where $\epsilon_{}=t/(P+Q)$.
With $v={\Delta x/{\Delta t}}=x/t={{(P-Q)}/{(P+Q)}}$ the classical
velocity attributed to the particle, $PQ=((P+Q)/2\gamma)^2$ where 
$\gamma=1/\sqrt{1-v^2}$. Thus we have
\begin{equation}
\Psi_{-+}= \sum^{\infty}_{k=0}(-1)^k {({m t/{2\gamma}})^{2k}\over{[(k)!]^2}}
= J_0(m t/\gamma)
\end{equation}
where $J_0$ is the zeroth order Bessel function of the first kind.
A similar calculation yields for $\Psi_{+-}$ the same result.

For $\Psi_{++}$, the number of bends to the right and to the left is $R/2$ 
for each direction where $R$ is even. However, the path is again defined
by $R^+=R/2$ bends to the right and $R^-=R/2-1$ bends to the left. Thus,
\begin{eqnarray}
\Psi_{++}
&=& \sum_{\mbox{\tiny even } R} (\mbox{i} m \epsilon_{})^{R-1} {P^{R/2} Q^{R/2-1}\over{(R/2)!(R/2-1)!}} \nonumber \\
&=& \mbox{i}\sqrt{{P/{Q}}}\sum^{\infty}_{k=0} (-1)^k {(m t/2\gamma)^{2k+1}\over{(k+1)!(k)!}} \nonumber \\
&=& \mbox{i}\sqrt{{P/{Q}}} J_1(m t/\gamma) \; .
\end{eqnarray}
With $\sqrt{{P/{Q}}}= (t+x)/(t^2-x^2)^{1/2}$ and $\tau=(t^2-x^2)^{1/2}$ the 
component $\Psi_{++}$ becomes
\begin{equation}
\Psi_{++}=\mbox{i} (t+x)/\tau \, J_1(m t/\gamma) \, .
\end{equation} 
A similar calculation yields 
\begin{equation}
\Psi_{--}=\mbox{i} (t-x)/\tau \, J_1(m t/\gamma) \, .
\end{equation} 
This completes the envisaged computation. As a side remark note that the presented 
calculation scheme is not restricted to $\epsilon_{j_r} \doteq \epsilon$. As will 
be shown elsewhere, similar results may be obtained for 
$\epsilon_{j_r} = \epsilon(j_r)$.

\section{Discussion}
To relate the components $\Psi_{\delta\gamma}$ to the solution of the 
Dirac equation (\ref{diracequ}) consider the explicit represation
\begin{equation}
\sigma_x = 
\left( \begin{array}{cc} 
  0 & 1 \\
  1 & 0
\end{array} \right), \quad
\sigma_z = 
\left( \begin{array}{cc} 
  1 & 0 \\
  0 & -1
\end{array} \right) \; . 
\end{equation}
As may be seen by direct calculation, in this representation 
$\Psi_1$ and $\Psi_2$ defined as 
\begin{equation}
\Psi_1=\left( \begin{array}{c} 
   \Psi_{++} \\
   \Psi_{+-}
\end{array} \right); \quad
\Psi_2=\left( \begin{array}{c} 
   \Psi_{+-} \\
   \Psi_{--}
\end{array} \right)
\end{equation}
are two independent, {\it exact} solutions of the Dirac equation (\ref{diracequ}).
This completes the demonstration that Feynman's chessboard model yields exact 
solutions to the Dirac equation when taking into account only those bends 
which {\it actually} define paths. 
With regard to fundamental theories of 
spacetime and/or quantum mechanics (e.g. in the spirit of Finkelstein, 1974) 
this could be of importance. Somehow similar results have been 
obtained from the continuum limit of a discretized version of the Dirac 
equation (Kauffman and Noyes, 1996). 

The calculation scheme and part of the results presented here can be generalized 
to unevenly spaced spacetime lattices. This opens up the possibility to define an analogon 
to the Feynman checkerboard for discrete spacetime models of different type (e.g. Kull 
and Treumann, 1994). Related work is in progress and will be presented elsewhere.

A.K. would like to thank Dr. O. Forster for valuable discussions. Part of the work of A.K.
has been supported by the Swiss National Foundation, grant 81AN-052101.

\pagebreak

\end{document}